\pdfoutput=1

\documentclass[12pt,a4paper]{article}

\usepackage{ifthen} 
\newboolean{pdflatex}
\setboolean{pdflatex}{true} 

\newboolean{articletitles}
\setboolean{articletitles}{true} 

\newboolean{uprightparticles}
\setboolean{uprightparticles}{false} 

\def\paperauthors{LHCb collaboration} 
\def\paperasciititle{Model-independent study of structure in B+ -> D+ D- K+ decays} 
\def\papertitle{Model-independent study of structure in $\Bp\to\Dp\Dm\Kp$ decays} 
\def\paperkeywords{{High Energy Physics}, {LHCb}} 
\def\papercopyright{\the\year\ CERN for the benefit of the LHCb collaboration} 
\def\paperlicence{CC BY 4.0 licence}
\def\paperlicenceurl{https://creativecommons.org/licenses/by/4.0/}

\usepackage[top=1in, bottom=1.25in, left=1in, right=1in]{geometry}

\usepackage{microtype}
\usepackage{lineno}  
\usepackage{xspace} 
\usepackage{caption} 

\usepackage{graphicx}  
\usepackage{color}
\usepackage{colortbl}
\graphicspath{{./figs/}} 

\usepackage{amsmath} 
\usepackage{amssymb}
\usepackage{amsfonts}
\usepackage{upgreek} 

\newcommand*\patchAmsMathEnvironmentForLineno[1]{%
\expandafter\let\csname old#1\expandafter\endcsname\csname #1\endcsname
\expandafter\let\csname oldend#1\expandafter\endcsname\csname
end#1\endcsname
 \renewenvironment{#1}%
   {\linenomath\csname old#1\endcsname}%
   {\csname oldend#1\endcsname\endlinenomath}%
}
\newcommand*\patchBothAmsMathEnvironmentsForLineno[1]{%
  \patchAmsMathEnvironmentForLineno{#1}%
  \patchAmsMathEnvironmentForLineno{#1*}%
}
\AtBeginDocument{%
\patchBothAmsMathEnvironmentsForLineno{equation}%
\patchBothAmsMathEnvironmentsForLineno{align}%
\patchBothAmsMathEnvironmentsForLineno{flalign}%
\patchBothAmsMathEnvironmentsForLineno{alignat}%
\patchBothAmsMathEnvironmentsForLineno{gather}%
\patchBothAmsMathEnvironmentsForLineno{multline}%
\patchBothAmsMathEnvironmentsForLineno{eqnarray}%
}

\usepackage{hyperxmp}

\usepackage[pdftex,
            pdfauthor={\paperauthors},
            pdftitle={\paperasciititle},
            pdfkeywords={\paperkeywords},
            pdfcopyright={Copyright (C) \papercopyright},
            pdflicenseurl={\paperlicenceurl}]{hyperref}

\usepackage[colorinlistoftodos,textsize=scriptsize]{todonotes}

\usepackage[bottom,flushmargin,hang,multiple]{footmisc}

\usepackage[all]{hypcap} 

\usepackage{xspace} 
\usepackage{upgreek}

\def\lhcb   {\mbox{LHCb}\xspace}

\def\MagUp {\mbox{\em Mag\kern -0.05em Up}\xspace}

\ifthenelse{\boolean{uprightparticles}}%
{

 \def\Pmu         {\ensuremath{\upmu}\xspace}

 \def\Ppi         {\ensuremath{\uppi}\xspace}

 \def\Ppsi        {\ensuremath{\uppsi}\xspace}

 \def\PDelta      {\ensuremath{\Delta}\xspace}                 
 \def\PXi         {\ensuremath{\Xi}\xspace}                 
 \def\PLambda     {\ensuremath{\Lambda}\xspace}                 
 \def\PSigma      {\ensuremath{\Sigma}\xspace}                 
 \def\POmega      {\ensuremath{\Omega}\xspace}                 
 \def\PUpsilon    {\ensuremath{\Upsilon}\xspace}

 \def\PB      {\ensuremath{\mathrm{B}}\xspace}                 
                  
 \def\PD      {\ensuremath{\mathrm{D}}\xspace}

 \def\PJ      {\ensuremath{\mathrm{J}}\xspace}                 
 \def\PK      {\ensuremath{\mathrm{K}}\xspace}

 \def\Pb      {\ensuremath{\mathrm{b}}\xspace}                 
 \def\Pc      {\ensuremath{\mathrm{c}}\xspace}

 \def\Pi      {\ensuremath{\mathrm{i}}\xspace}

 \def\Pp      {\ensuremath{\mathrm{p}}\xspace}

 \def\Ps      {\ensuremath{\mathrm{s}}\xspace}

 \def\thebaroffset{0.0em}
}
{

 \def\Pmu         {\ensuremath{\mu}\xspace}

 \def\Ppi         {\ensuremath{\pi}\xspace}

 \def\Ppsi        {\ensuremath{\psi}\xspace}                 
                  
 \mathchardef\PDelta="7101
 \mathchardef\PXi="7104
 \mathchardef\PLambda="7103
 \mathchardef\PSigma="7106
 \mathchardef\POmega="710A
 \mathchardef\PUpsilon="7107
                  
 \def\PB      {\ensuremath{B}\xspace}                 
                  
 \def\PD      {\ensuremath{D}\xspace}

 \def\PJ      {\ensuremath{J}\xspace}                 
 \def\PK      {\ensuremath{K}\xspace}

 \def\Pb      {\ensuremath{b}\xspace}                 
 \def\Pc      {\ensuremath{c}\xspace}

 \def\Pi      {\ensuremath{i}\xspace}

 \def\Pp      {\ensuremath{p}\xspace}

 \def\Ps      {\ensuremath{s}\xspace}

 \def\thebaroffset{0.18em}
}
\newcommand{\offsetoverline}[2][\thebaroffset]{\kern #1\overline{\kern -#1 #2}}%

\makeatletter
\ifcase \@ptsize \relax
  \newcommand{\miniscule}{\@setfontsize\miniscule{4}{5}}
\or
  \newcommand{\miniscule}{\@setfontsize\miniscule{5}{6}}
\or
  \newcommand{\miniscule}{\@setfontsize\miniscule{5}{6}}
\fi
\makeatother

\DeclareRobustCommand{\optbar}[1]{\shortstack{{\miniscule (\rule[.5ex]{1.25em}{.18mm})}
  \\ [-.7ex] $#1$}}

\def\mumu       {{\ensuremath{\Pmu^+\Pmu^-}}\xspace}

\def\squark    {{\ensuremath{\Ps}}\xspace}

\def\cquark    {{\ensuremath{\Pc}}\xspace}

\def\bquark    {{\ensuremath{\Pb}}\xspace}

\def\pion   {{\ensuremath{\Ppi}}\xspace}

\def\pip    {{\ensuremath{\pion^+}}\xspace}
\def\pim    {{\ensuremath{\pion^-}}\xspace}

\def\kaon    {{\ensuremath{\PK}}\xspace}

\def\KorKbar {\kern \thebaroffset\optbar{\kern -\thebaroffset \PK}{}\xspace}

\def\Kp      {{\ensuremath{\kaon^+}}\xspace}
\def\Km      {{\ensuremath{\kaon^-}}\xspace}

\def\D       {{\ensuremath{\PD}}\xspace}

\def\DorDbar {\kern \thebaroffset\optbar{\kern -\thebaroffset \PD}\xspace}

\def\Dp      {{\ensuremath{\D^+}}\xspace}
\def\Dm      {{\ensuremath{\D^-}}\xspace}
\def\Dpm     {{\ensuremath{\D^\pm}}\xspace}

\def\DpDm    {\ensuremath{\Dp {\kern -0.16em \Dm}}\xspace}

\def\B       {{\ensuremath{\PB}}\xspace}

\def\BorBbar {\kern \thebaroffset\optbar{\kern -\thebaroffset \PB}\xspace}

\def\Bd      {{\ensuremath{\B^0}}\xspace}

\def\BdorBdbar {\kern \thebaroffset\optbar{\kern -\thebaroffset \Bd}\xspace}
\def\Bu      {{\ensuremath{\B^+}}\xspace}

\def\Bp      {{\ensuremath{\Bu}}\xspace}

\def\Bs      {{\ensuremath{\B^0_\squark}}\xspace}

\def\BsorBsbar {\kern \thebaroffset\optbar{\kern -\thebaroffset \Bs}\xspace}

\def\jpsi     {{\ensuremath{{\PJ\mskip -3mu/\mskip -2mu\Ppsi}}}\xspace}

\def\Y#1S{\ensuremath{\PUpsilon{(#1S)}}\xspace}

\def\proton      {{\ensuremath{\Pp}}\xspace}

\def\Lz          {{\ensuremath{\PLambda}}\xspace}

\def\LorLbar     {\kern \thebaroffset\optbar{\kern -\thebaroffset \PLambda}\xspace}

\def\Lb           {{\ensuremath{\Lz^0_\bquark}}\xspace}

\def\to                 {\ensuremath{\rightarrow}\xspace}

\def\AT#1     {\ensuremath{A_{\mathrm{T}}^{#1}}\xspace}           

\def\C#1      {\ensuremath{\mathcal{C}_{#1}}\xspace}                       
\def\Cp#1     {\ensuremath{\mathcal{C}_{#1}^{'}}\xspace}                    
\def\Ceff#1   {\ensuremath{\mathcal{C}_{#1}^{\mathrm{(eff)}}}\xspace}        
\def\Cpeff#1  {\ensuremath{\mathcal{C}_{#1}^{'\mathrm{(eff)}}}\xspace}       
\def\Ope#1    {\ensuremath{\mathcal{O}_{#1}}\xspace}                       
\def\Opep#1   {\ensuremath{\mathcal{O}_{#1}^{'}}\xspace}                    

\newcommand{\aunit}[1]{\ensuremath{\text{\,#1}}}       

\newcommand{\tev}{\aunit{Te\kern -0.1em V}\xspace}
\newcommand{\gev}{\aunit{Ge\kern -0.1em V}\xspace}
\newcommand{\mev}{\aunit{Me\kern -0.1em V}\xspace}
\newcommand{\kev}{\aunit{ke\kern -0.1em V}\xspace}
\newcommand{\ev}{\aunit{e\kern -0.1em V}\xspace}
 
\newcommand{\mevc}{\ensuremath{\aunit{Me\kern -0.1em V\!/}c}\xspace}
\newcommand{\gevc}{\ensuremath{\aunit{Ge\kern -0.1em V\!/}c}\xspace}
\newcommand{\mevcc}{\ensuremath{\aunit{Me\kern -0.1em V\!/}c^2}\xspace}
\newcommand{\gevcc}{\ensuremath{\aunit{Ge\kern -0.1em V\!/}c^2}\xspace}

\def\fb   {\ensuremath{\aunit{fb}}\xspace}
\def\invfb   {\ensuremath{\fb^{-1}}\xspace}

\def\gsim{{~\raise.15em\hbox{$>$}\kern-.85em
          \lower.35em\hbox{$\sim$}~}\xspace}
\def\lsim{{~\raise.15em\hbox{$<$}\kern-.85em
          \lower.35em\hbox{$\sim$}~}\xspace}

\def\tell1  {TELL1\xspace}
\def\ukl1   {UKL1\xspace}

\newcommand{\ie}{\mbox{\itshape i.e.}\xspace}

\usepackage{cite} 
\usepackage{mciteplus}

\usepackage{listings}
\def\BpDpDmKp{$\Bp\to\Dp\Dm\Kp$\xspace}

\begin{document}

\renewcommand{\thefootnote}{\fnsymbol{footnote}}
\setcounter{footnote}{1}

\begin{titlepage}
\pagenumbering{roman}

\vspace*{-1.5cm}
\centerline{\large EUROPEAN ORGANIZATION FOR NUCLEAR RESEARCH (CERN)}
\vspace*{1.5cm}
\noindent
\begin{tabular*}{\linewidth}{lc@{\extracolsep{\fill}}r@{\extracolsep{0pt}}}
\ifthenelse{\boolean{pdflatex}}
{\vspace*{-1.5cm}\mbox{\!\!\!\includegraphics[width=.14\textwidth]{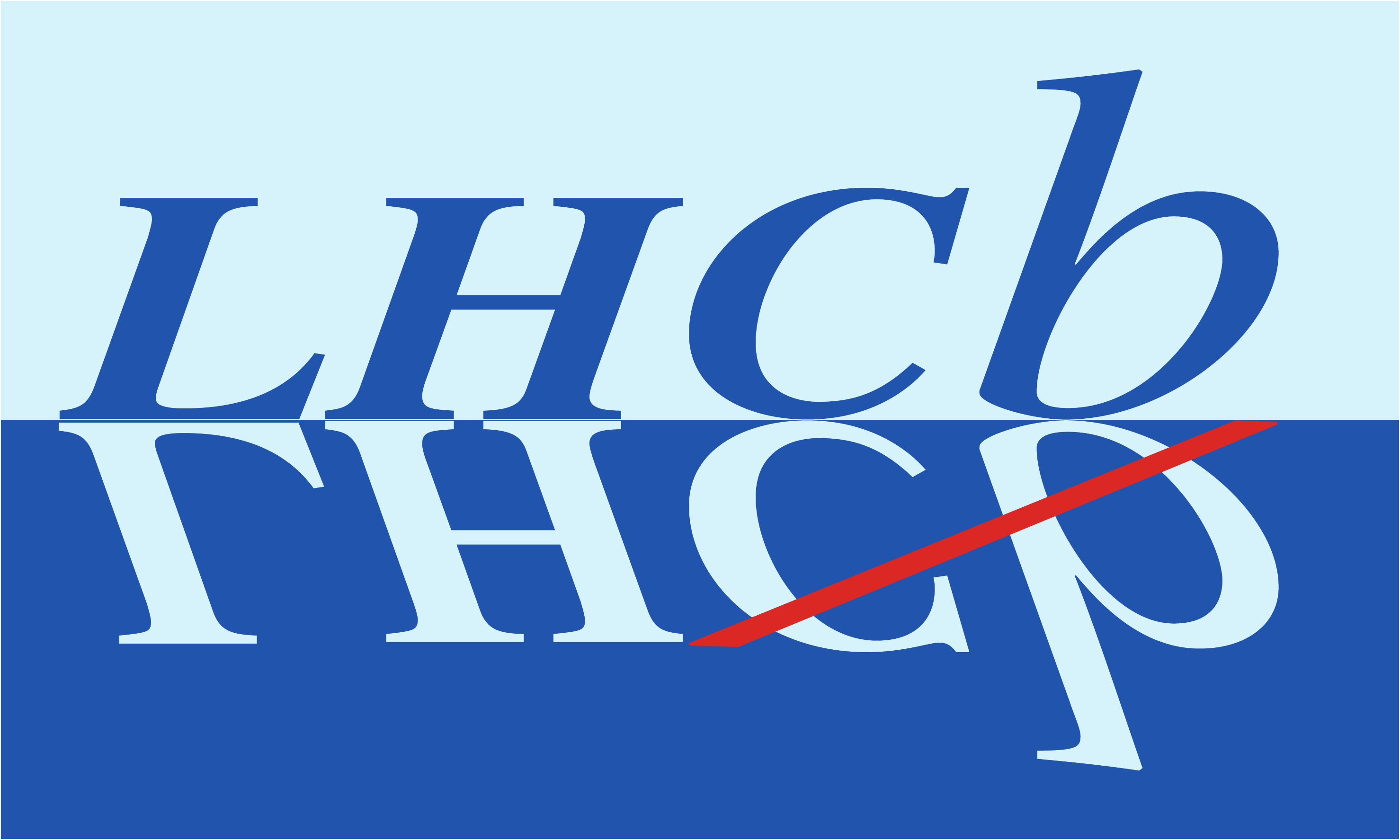}} & &}%
{\vspace*{-1.2cm}\mbox{\!\!\!\includegraphics[width=.12\textwidth]{figs/lhcb-logo.eps}} & &}%
\\
 & & CERN-EP-2020-158 \\  
 & & LHCb-PAPER-2020-024 \\  
 & & December 7, 2020 \\ 
 & & \\
\end{tabular*}

\vspace*{4.0cm}

{\normalfont\bfseries\boldmath\huge
\begin{center}
  \papertitle 
\end{center}
}

\vspace*{2.0cm}

\begin{center}
\paperauthors\footnote{Authors are listed at the end of this Letter.}
\end{center}

\vspace{\fill}

\begin{abstract}
  \noindent
The only anticipated resonant contributions to \BpDpDmKp\ decays are charmonium states in the $\Dp\Dm$ channel. 
A model-independent analysis, using \lhcb proton-proton collision data taken at centre-of-mass energies of $\sqrt{s}=7,8,$ and $13\tev$, corresponding to a total integrated luminosity of 9\invfb, is carried out to test this hypothesis.
The description of the data assuming that resonances only manifest in decays to the $\Dp\Dm$ pair is shown to be incomplete. 
This constitutes evidence for a new contribution to the decay, potentially one or more new charm-strange resonances in the $\Dm\Kp$ channel with masses around 2.9\gevcc.
\end{abstract}

\vspace*{2.0cm}

\begin{center}
  Published in 
  Phys.~Rev.~Lett. 125 (2020) 242001
\end{center}

\vspace{\fill}

{\footnotesize 
\centerline{\copyright~\papercopyright. \href{\paperlicenceurl}{\paperlicence}.}}
\vspace*{2mm}

\end{titlepage}


\newpage
\setcounter{page}{2}
\mbox{~}
%
%
%
%


\renewcommand{\thefootnote}{\arabic{footnote}}
\setcounter{footnote}{0}
\cleardoublepage

\pagestyle{plain} 
\setcounter{page}{1}
\pagenumbering{arabic}



The $\Bp\to D{}^{(*)+}D{}^{(*)-}\Kp$ family of decays offers unique opportunities to study charmonium states.
The constrained environment of $B$ meson decays allows the masses, widths and quantum numbers of such states to be determined using amplitude analysis techniques, with low backgrounds from other processes. 
In particular, resonances in the $D{}^{(*)-}\Kp$ or $D{}^{(*)+}\Kp$ channels would be manifestly exotic, having minimal quark content $\bar{c}du\bar{s}$ or $c\bar{d}u\bar{s}$, respectively.
While many exotic hadrons containing $c\bar{c}$ or $b\bar{b}$ quarks have recently been observed~\cite{Karliner:2017qhf,Olsen:2017bmm,Ali:2017jda}, there is to date no significant evidence of the existence of exotic hadrons with open flavour, \ie\ with non-zero strangeness, charm, or beauty quantum numbers. 
Studies of $\Bp\to D{}^{(*)+}D{}^{(*)-}\Kp$ decays are therefore expected to help resolve open questions regarding charmonium spectroscopy~\cite{X3915_chic02P,Duan:2020tsx}.
In addition, measurements of these processes have been proposed as a method to aid characterisation of the $c\bar{c}$ contribution in $\Bp\to\Kp\mumu$ decays~\cite{Khodjamirian:2010vf,Lyon:2014hpa}.

The branching fractions of $\Bp\to D{}^{(*)+}D{}^{(*)-}\Kp$ decays have been measured~\cite{delAmoSanchez:2010pg,LHCb-PAPER-2020-006}, but no prior analyses of their resonant structure exist.\footnote{The inclusion of charge-conjugate processes is implied throughout this Letter.}
Recent studies have shown that extremely pure samples of these decays can be obtained using LHCb data~\cite{LHCb-PAPER-2020-006}, with yields much larger than those available at previous experiments.

A model-dependent study of the resonant structure in \BpDpDmKp decays~\cite{LHCb-PAPER-2020-025}, carried out in parallel to this work, has revealed structure in the $\Dm\Kp$ invariant-mass spectrum that cannot be described by reflections of charmonium resonances. This highly surprising observation, along with the limited current knowledge of the charmonium spectrum in this mass range, particularly among spin-0 and spin-2 states, motivates the study of this decay using a model-independent approach as presented in this Letter. This method is particularly useful when applied to three-body decays where resonances are only expected to form between one pair of the final-state particles, such that the decay kinematics are described through one mass and one angular variable. Unexpected, exotic, contributions to the decay process manifest as high-order moments in the distribution of the angular variable, as has been demonstrated by the use of the method to identify exotic resonances contributing to $\Bd\to\psi(2S)\Kp\pim$~\cite{LHCb-PAPER-2015-038}, $\Lb\to\jpsi\proton\Km$~\cite{LHCb-PAPER-2016-009} and $\Bd\to\jpsi\Kp\pim$~\cite{LHCb-PAPER-2018-043} decays.

The model-independent analysis of the \BpDpDmKp\ decay involves consideration of the distribution of the variable $h(\Dp\Dm)$ defined as the cosine of the $\Dp\Dm$ helicity angle, \ie\ the angle between the momenta of the \Kp and \Dm particles in the \Dp\Dm rest frame.
A description of the \BpDpDmKp Dalitz plot is obtained by decomposing the $h(\Dp\Dm)$ distribution in terms of Legendre polynomials.
The decomposition is done within slices of the \Dp\Dm invariant mass, $m(\Dp\Dm)$, thereby accounting for the two degrees of freedom in the \BpDpDmKp decay kinematics. 
The description can be projected onto the other invariant-mass distributions in order to identify regions where exotic contributions are needed, and the significance of such deviations can be quantified. If only $\Dp\Dm$ resonances contribute, the projections will be well described using only low-order moments, up to twice the maximum spin of charmonium resonances present. If peaking contributions from other channels enter, higher-order moments will be required. The narrower the structure, the higher the order that will be needed.
Consequently, a description employing only low-order moments will be incomplete.


This method is applied to a sample of \BpDpDmKp candidates selected from \lhcb proton-proton ($pp$) collision datasets, corresponding to integrated luminosities of 3\invfb recorded during 2011 and 2012 (Run 1) and 6\invfb from 2015 to 2018 (Run 2). 
The data sample, selection criteria, background and efficiency modelling are identical to those in the amplitude analysis of the same process, described in detail in Ref.~\cite{LHCb-PAPER-2020-025} and briefly summarised here.
The LHCb detector~\cite{LHCb-DP-2008-001,LHCb-DP-2014-002} is a single-arm forward spectrometer covering the pseudorapidity range $2 < \eta < 5$, designed for the study of particles containing \bquark\ or \cquark\ quarks. 
Simulation, produced with software packages described in Refs.~\cite{Sjostrand:2007gs,*Sjostrand:2006za,Lange:2001uf,Allison:2006ve,*Agostinelli:2002hh,LHCb-DP-2018-004}, is used to model the effects of the detector acceptance and the imposed selection requirements.
The online event selection is performed by a trigger~\cite{LHCb-DP-2012-004}, which consists of a hardware stage, based on information from the calorimeter and muon systems, followed by a software stage, which applies a full event reconstruction and which identifies a two-, three-, or four-track secondary vertex by means of a multivariate algorithm. 
The charm mesons are reconstructed via the $\Dp\to\Km\pip\pip$ decay. Reconstructed \BpDpDmKp candidates that pass the trigger criteria are subjected to further selection requirements, including the use of a boosted decision tree (BDT) algorithm~\cite{Breiman,Stevens:2013dya} to reduce combinatorial background. Variables characterising the particular topology of the decay (flight distance of the $D$ mesons and displacement of the reconstructed intermediate and final-state particles from the $B$ production point) and particle identification information are used as inputs to the BDT algorithm.
Specific requirements are imposed to suppress contributions from \B decays involving one or no \D mesons, but having the same set of final-state pions and kaons as the signal decays; inspection of the sidebands of the \D-candidates' invariant-mass distributions confirms that any residual background from this source is at a negligible level.

\begin{figure}[!tb]
\centering
\includegraphics[width=.49\textwidth]{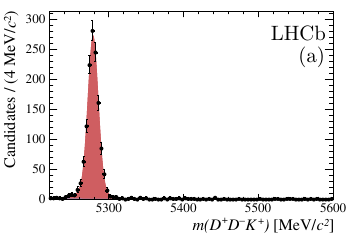}
\includegraphics[width=.49\textwidth]{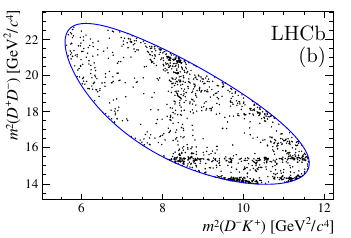}
\caption{(a)~Invariant-mass distribution for \B candidates with the results of the fit superimposed, where the signal component is indicated in red and background (barely visible) in blue. (b)~Dalitz plot for candidates with $m(\Dp\Dm\Kp)$ values in the signal window.\label{im_massfit_and_dalitz}}
\end{figure}

An extended maximum-likelihood fit is applied to the invariant-mass, $m(\Dp\Dm\Kp)$, distribution of the selected candidates, shown in Fig.~\ref{im_massfit_and_dalitz}(a).
There are 1260 candidates inside the signal window of $m(\Dp\Dm\Kp)$ within $20\mevcc$ of the known \Bp mass~\cite{PDG2019}, in which the sample purity is greater than 99.5\% and the residual background is combinatorial in nature.
The distribution of these candidates, which are retained for further analysis, in the Dalitz plot is shown in Fig.~\ref{im_massfit_and_dalitz}(b). 
The Dalitz-plot coordinates, $m^2(\Dm\Kp)$ and $m^2(\Dp\Dm)$, are determined after refitting the candidate decays, imposing the constraints that the reconstructed \Bp and \Dpm masses should match their known values and that the reconstructed \Bp meson should originate at its associated primary $pp$ interaction vertex. 
Charmonium resonances are clearly visible as horizontal bands in the Dalitz plot, but additional structure also appears to be present. 
A signal efficiency map is determined as a function of position in the Dalitz plot with simulation, where the particle identification response is calibrated using data control samples~\cite{LHCb-DP-2018-001,Poluektov:2014rxa}.  
The efficiency is found to vary with $m(\Dp\Dm)$ at the $\pm 10\%$ level, and to depend only weakly on $h(\Dp\Dm)$.

The $m(\Dp\Dm)$ distribution is divided into slices of width 20\mevcc, which is large compared to the resolution but narrower than any expected structure. Within each slice the distribution of the cosine of the helicity angle is decomposed according to the basis of Legendre polynomials. Including a factor to ensure normalisation over the domain \mbox{$-1$ to $1$}, these are given by
\begin{equation}
P_n\left(h(\Dp\Dm)\right) = \sqrt{\frac{2n+1}{2}} \times  2^n \sum_{r=0}^n \left( h(\Dp\Dm) \right)^r {n \choose r} {\frac{n + r -1}{2} \choose n}\,.
\end{equation}
In bin $j$ of the $m(\Dp\Dm)$ distribution, the coefficient of the expansion at order $k$ is referred to as the $k$-th unnormalised moment,
\begin{equation}
\langle Y^{j}_k \rangle = \sum_{l=1}^{N^{\rm Data}_{j}} w_l P_k\left(h_l(\Dp\Dm)\right)\,,
\label{momentsDefinition}
\end{equation}
where the sum is over the $N^{\rm Data}_{j}$ candidates in that bin, $w_l$ is a weight assigned to each candidate to achieve a background subtraction and efficiency correction, and $h_l(\Dp\Dm)$ is the value of $h(\Dp\Dm)$ for candidate $l$.
To probe whether charmonium resonances with spins up to and including $J_{\rm max}$ account for the structures observed in the Dalitz plot, the expansion can be truncated at a given order, $k_{\rm max} = 2J_{\rm max}$.

A simulated sample, generated uniformly in the Dalitz plot and weighted using the truncated expansion, is used in order to visualise the description of the $m(\Dm\Kp)$ and $m(\Dp\Kp)$ distributions and to compare them to data. 
The weights applied to the simulated sample are
\begin{equation}
\eta_i = \frac{2}{N^{\rm Sim}_j} \times  \sum_{k=0}^{k_{\rm max}} \langle Y^{j}_k\rangle P_k\left(h_i(\Dp\Dm)\right)\,,
\label{eqn_momentsWeightMC}
\end{equation}
where $i$ indexes the generated candidates and $N^{\rm Sim}_j$ is the number of candidates in the simulation in bin $j$ of the $m(\Dp\Dm)$ spectrum, centred on $m_j(\Dp\Dm)$. 

The significance of any deviation between the truncated Legendre polynomial description and the data can be assessed using pseudoexperiments. They are generated according to a probability density function (PDF)  constructed as a function of $m(\Dp\Dm)$ and $h(\Dp\Dm)$, given an hypothesis $H$ regarding $k_{\rm max}$,
\begin{equation}
    \mathcal{P}\left(m_j(\Dp\Dm),h(\Dp\Dm) | H\right) = \mathcal{P}\left(m_j(\Dp\Dm) \right) \mathcal{P}\left(h(\Dp\Dm) | H,m_j(\Dp\Dm)\right)\,.
\label{eqn_twodweighting}
\end{equation}
The binned PDF $\mathcal{P}\left(m_j(\Dp\Dm) \right)$ is given by
\begin{equation}
    \mathcal{P}\left(m_j(\Dp\Dm) \right) = \mathcal{N} \sum_{l=1}^{N_j^{\rm Data}} w_l\,,
\end{equation}
where $\mathcal{N}$ is a normalisation factor. The PDF $\mathcal{P}(h(\Dp\Dm) | H,m_j(\Dp\Dm))$ is a function of the moments and Legendre polynomial functions, reproducing the helicity angle dependence in bin $j$ of $m(\Dp\Dm)$,
\begin{equation}
    \mathcal{P}\left(h(\Dp\Dm) | H,m_j(\Dp\Dm)\right) = 1 + \frac{2}{\sum_{l=1}^{N^{\rm Data}_j}\!\!\!w_l} \, \sum_{k=1}^{k_{\rm max}} \langle Y^{j}_k\rangle P_k\left(h(\Dp\Dm)\right)\,.
    \label{expr_MIPDFDalitz}
\end{equation}
Since reflections of exotic contributions to the $\Dm\Kp$ or $\Dp\Kp$ channels would produce complicated structure in the $\left( m(\Dp\Dm), h(\Dp\Dm) \right)$ plane, the most sensitive model-independent test statistic is based on the PDF for $m(\Dm\Kp)$ or $m(\Dp\Kp)$. The PDF $\mathcal{P}\left(h(\Dp\Dm) | H, m_j(\Dp\Dm)\right)$ is projected onto $m(\Dm\Kp)$ or $m(\Dp\Kp)$ by generating candidates uniformly in the $\left( m(\Dp\Dm), h(\Dp\Dm) \right)$ plane and assigning a weight to each according to Eq.~\eqref{eqn_twodweighting}. A representation of $\mathcal{P}\left(m(\Dm\Kp) | H\right)$ or $\mathcal{P}\left(m(\Dp\Kp) | H\right)$ is then obtained by filling a histogram of $m(\Dm\Kp)$ or $m(\Dp\Kp)$ with these weighted candidates, respectively. 

A test statistic is constructed to discriminate between the hypothesis, $H_0$, that only $\Dp\Dm$ resonances contribute up to order $k_{\rm max}$ and the hypothesis that allows for contributions from higher-order moments to describe higher-spin or exotic contributions, $H_1$. The test statistic, formulated in terms of determining the significance of deviations in the $\Dm\Kp$ channel, has the form~\cite{LHCb-PAPER-2015-029}
\begin{equation}
    t = -2 \sum_{l=1}^{N^{\rm Data}} s_l \log\left( \frac{\mathcal{P}\left(m_l(\Dm\Kp) | H_0\right)/I_{H_0}}{\mathcal{P}\left(m_l(\Dm\Kp) | H_1\right)/I_{H_1}} \right)\,,
    \label{expr_testStatistic}
\end{equation}
where $\mathcal{P}\left(m_l(\Dm\Kp) | H\right)$ is the value of the PDF in the bin of $m(\Dm\Kp)$ where candidate $l$ is found, $s_l$ is the signal weight effecting a background subtraction~\cite{Pivk:2004ty}, and $I_{H}$ is a normalisation factor, computed by Monte Carlo integration,
\begin{equation}
    I_{H} = \sum_{l=1}^{N^{\rm Sim}} \mathcal{P}(m_l(\Dm\Kp) | H) \epsilon_l\,,
\end{equation}
where $\epsilon_l$ is the efficiency appropriate for candidate $l$.

\begin{figure}[!tb]
\centering
\includegraphics[width=.995\textwidth]{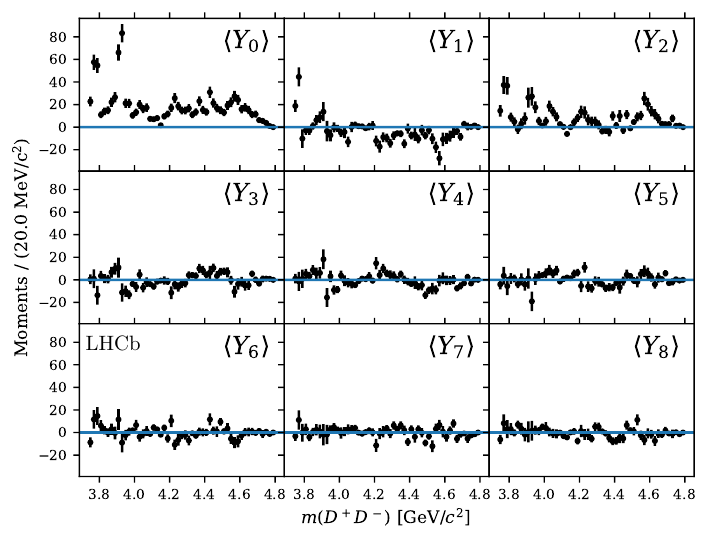}
\caption{Distributions of the first nine unnormalised moments, $\langle Y_k^j\rangle$, defined in Eq.~\eqref{momentsDefinition}, as a function of $m(\Dp\Dm)$ for the selected \BpDpDmKp candidates, after efficiency correction and background subtraction have been applied.\label{im_unnorm_moments_m12}}
\end{figure}

The distributions in the $\Dp\Dm$ invariant mass, $m(\Dp\Dm)$, of the first nine unnormalised moments defined in Eq.~\eqref{momentsDefinition} are computed for the selected candidates, and are shown in Fig.~\ref{im_unnorm_moments_m12}. 
Significant structure is visible at $m(\Dp\Dm) \approx 3.8\gevcc$ up to and including the second moment, and not at higher orders, as expected for a contribution from the spin-1 resonance $\psi(3770)$. In the vicinity of the $\chi_{c2}(2P)$ resonance near $m(\Dp\Dm)=3.9\gevcc$, significant structure appears at order two and persists, albeit weakly, at order four. 
This is found, in the model-dependent analysis~\cite{LHCb-PAPER-2020-025}, to be due to the presence of both spin-0 and spin-2 charmonia in this region.
Structure at low $m(\Dp\Dm)$ in the first moment indicates interference between S~and P~waves and, similarly, that around $m(\Dp\Dm)=3.9\gevcc$ in the third moment could indicate interference between P~and D~waves. Structure apparent at all orders for $m(\Dp\Dm)>4.1\gevcc$ --- though having large uncertainties at orders above 5 --- could indicate reflection from a structure in another two-body combination. 

\begin{figure}[!tb]
\centering
\includegraphics[width=.48\textwidth]{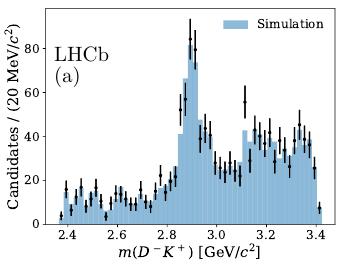}
\includegraphics[width=.48\textwidth]{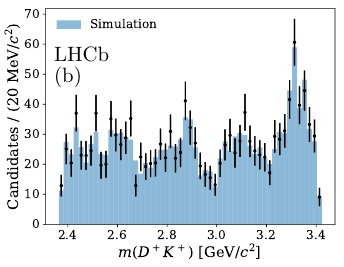}
\caption{Comparison between data (points with error bars) and a weighted generated sample (filled histogram) as a function of (a) $m(\Dm\Kp)$ and (b) $m(\Dp\Kp)$, where the weights account for the Legendre polynomial moments of order up to and including 29.
\label{im_dataMC_agreement_highOrder}}
\end{figure}

In order to test how well the \BpDpDmKp Dalitz plot can be described using a truncated sum over $m(\Dp\Dm)$ moments, a sample of $10^7$ \BpDpDmKp decays is generated uniformly in the $\left( m(\Dp\Dm), h(\Dp\Dm) \right)$ plane. Weights are applied according to Eq.~\eqref{eqn_momentsWeightMC}, and the resulting distribution of the weighted sample is compared to that for the candidates selected from the \lhcb data. In the first instance, $k_{\rm max}$ is set to a high value of 29 in the construction of weights, to allow all but the smallest of fluctuations in data to be captured. The comparison between the generated decays and the data sample is shown in Fig.~\ref{im_dataMC_agreement_highOrder}. 
The excellent agreement, limited only by statistical fluctuations which can generate structure to arbitrarily high moments, in the $m(\Dm\Kp)$ and $m(\Dp\Kp)$  invariant-mass distributions is also to be expected, given the high value of $k_{\rm max}$.

\begin{figure}[!tb]
\centering
	\includegraphics[width=.48\textwidth]{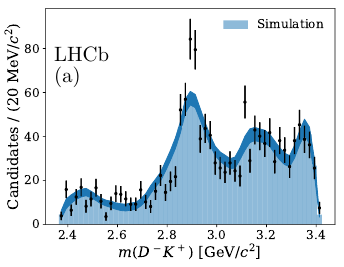}
    \includegraphics[width=.48\textwidth]{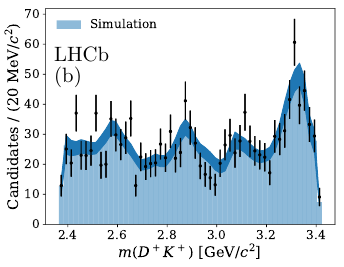}
\caption{Comparison between data (points with error bars) and a weighted generated sample (filled histogram) as a function of (a) $m(\Dm\Kp)$ and (b) $m(\Dp\Kp)$, where the weights account for the Legendre polynomial moments of order up to and including four. The uncertainty on the weighted shape (dark band) is also shown.\label{im_dataMC_agreement_lowOrder}}
\end{figure}

The effect of truncating the sum over moments at a lower value is explored. A value of $k_{\rm max}=4$ is chosen under the assumption that only resonances with spin up to 2 appear in the $\Dp\Dm$ channel, since production of high-spin resonances in \B-meson decays is suppressed and no evidence for a contribution with spin-3 or higher is seen in either Fig.~\ref{im_unnorm_moments_m12} or the model-dependent analysis~\cite{LHCb-PAPER-2020-025}. 
Figure~\ref{im_dataMC_agreement_lowOrder} shows the comparison between the weighted generated sample and the data. A prominent discrepancy is apparent around $m(\Dm\Kp)=2.9\gevcc$.
No narrow regions of disagreement are evident in the \Dp\Kp spectrum.

\begin{figure}[!tb]
\centering
\includegraphics[width=.48\textwidth]{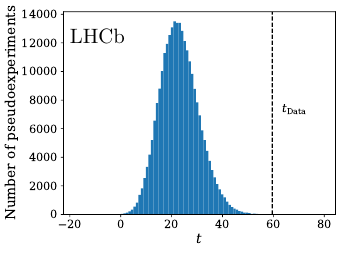}
\caption{Comparison of the test statistic evaluated for the data (black dashed line) and for the ensemble of pseudoexperiments (blue histogram) generated according to the PDF constructed using the first four moments of the $h(\Dp\Dm)$ distribution in data.\label{im_significance}}
\end{figure}

The significance of the discrepancy in the $m(\Dm\Kp)$ distribution between the data and the weighted generated sample in Fig.~\ref{im_dataMC_agreement_lowOrder}(a) is evaluated using the test statistic defined in Eq.~\eqref{expr_testStatistic}. An ensemble of pseudoexperiments, in which each dataset has the same size as the real dataset, is prepared according to the PDF defined in Eq.~\eqref{expr_MIPDFDalitz}, where $k_{\rm max}$ is taken to be 4. 
The tiny background contribution is ignored, which introduces negligible uncertainty due to the high purity of the selected \BpDpDmKp\ sample.
For each pseudoexperiment, a new efficiency map is generated to incorporate the systematic uncertainty arising from the limited size of the simulated sample. 
This ensemble of nearly 260\,000 pseudoexperiments allows determination of the distribution of the test statistic under the hypothesis, $H_0$, that only $\Dp\Dm$ resonances up to spin-2 are present, as shown in Fig.~\ref{im_significance}. 
The value of the test statistic obtained from data, $t_{\rm Data}$, allows the $H_0$ hypothesis to be rejected at the 99.994\% level, corresponding to a significance of 3.9 Gaussian standard deviations ($\sigma$). Even when moments up to order 6 are considered, the significance of the discrepancy remains above 3.7$\sigma$.

In summary, a model-independent technique has been employed to confirm whether or not the observed $m(\Dm\Kp)$ distribution in \BpDpDmKp decays reconstructed in the  LHCb data sample can be explained in terms of reflections from charmonium resonances alone. 
It is found that the intermediate structure of the decay cannot be described using only \Dp\Dm resonances of spin up to 2.
The significance of the disagreement in the $m(\Dm\Kp)$ distribution is $3.9\,\sigma$, and is most apparent in the region $m(\Dm\Kp)=2.9\gevcc$. 
This discrepancy could be explained by a new, manifestly exotic, charm-strange resonance decaying to the $\Dm\Kp$ final state.
The outcome of this model-independent study therefore supports the results of the amplitude analysis of the same data~\cite{LHCb-PAPER-2020-025}, where both spin-0 and spin-1 components are included in the $\Dm\Kp$ channel, as well as $\psi(3770)$, $\chi_{c0}(3930)$, $\chi_{c2}(3930)$, $\psi(4040)$, $\psi(4160)$, and $\psi(4415)$ resonances decaying to $\Dp\Dm$, and a nonresonant component.

\section*{Acknowledgements}
%
%
\noindent We express our gratitude to our colleagues in the CERN
accelerator departments for the excellent performance of the LHC. We
thank the technical and administrative staff at the LHCb
institutes.
We acknowledge support from CERN and from the national agencies:
CAPES, CNPq, FAPERJ and FINEP (Brazil); 
MOST and NSFC (China); 
CNRS/IN2P3 (France); 
BMBF, DFG and MPG (Germany); 
INFN (Italy); 
NWO (Netherlands); 
MNiSW and NCN (Poland); 
MEN/IFA (Romania); 
MSHE (Russia); 
MICINN (Spain); 
SNSF and SER (Switzerland); 
NASU (Ukraine); 
STFC (United Kingdom); 
DOE NP and NSF (USA).
We acknowledge the computing resources that are provided by CERN, IN2P3
(France), KIT and DESY (Germany), INFN (Italy), SURF (Netherlands),
PIC (Spain), GridPP (United Kingdom), RRCKI and Yandex
LLC (Russia), CSCS (Switzerland), IFIN-HH (Romania), CBPF (Brazil),
PL-GRID (Poland) and OSC (USA).
We are indebted to the communities behind the multiple open-source
software packages on which we depend.
Individual groups or members have received support from
AvH Foundation (Germany);
EPLANET, Marie Sk\l{}odowska-Curie Actions and ERC (European Union);
A*MIDEX, ANR, Labex P2IO and OCEVU, and R\'{e}gion Auvergne-Rh\^{o}ne-Alpes (France);
Key Research Program of Frontier Sciences of CAS, CAS PIFI,
Thousand Talents Program, and Sci. \& Tech. Program of Guangzhou (China);
RFBR, RSF and Yandex LLC (Russia);
GVA, XuntaGal and GENCAT (Spain);
the Royal Society
and the Leverhulme Trust (United Kingdom).



\addcontentsline{toc}{section}{References}
\bibliographystyle{LHCb}
\bibliography{main,standard,LHCb-PAPER,LHCb-CONF,LHCb-DP,LHCb-TDR}


\newpage
\centerline
{\large\bf LHCb collaboration}
\begin
{flushleft}
\small
R.~Aaij$^{31}$,
C.~Abell{\'a}n~Beteta$^{49}$,
T.~Ackernley$^{59}$,
B.~Adeva$^{45}$,
M.~Adinolfi$^{53}$,
H.~Afsharnia$^{9}$,
C.A.~Aidala$^{84}$,
S.~Aiola$^{25}$,
Z.~Ajaltouni$^{9}$,
S.~Akar$^{64}$
% ...

{\footnotesize \it

% ...
$ ^{9}$Universit{\'e} Clermont Auvergne, CNRS/IN2P3, LPC, Clermont-Ferrand, France\\
% ...
$ ^{25}$INFN Sezione di Milano, Milano, Italy\\
% ...
$ ^{31}$Nikhef National Institute for Subatomic Physics, Amsterdam, Netherlands\\
% ...
$ ^{45}$Instituto Galego de F{\'\i}sica de Altas Enerx{\'\i}as (IGFAE), Universidade de Santiago de Compostela, Santiago de Compostela, Spain\\
% ...
$ ^{49}$Physik-Institut, Universit{\"a}t Z{\"u}rich, Z{\"u}rich, Switzerland\\
% ...
$ ^{53}$H.H. Wills Physics Laboratory, University of Bristol, Bristol, United Kingdom\\
% ...
$ ^{59}$Oliver Lodge Laboratory, University of Liverpool, Liverpool, United Kingdom\\
% ...
$ ^{64}$University of Cincinnati, Cincinnati, OH, United States\\
% ...
$ ^{67}$Syracuse University, Syracuse, NY, United States\\
% ...
$ ^{84}$University of Michigan, Ann Arbor, United States, associated to $^{67}$\\
\bigskip
% ...
\medskip
}
\end{flushleft}

\end{document}